\documentclass[sigconf]{acmart}
\AtBeginDocument{%
	\providecommand\BibTeX{{%
			\normalfont B\kern-0.5em{\scshape i\kern-0.25em b}\kern-0.8em\TeX}}}

\usepackage{amsmath,amsfonts,bm}

\def\eqref#1{equation~\ref{#1}}

\def\1{\bm{1}}

\def\va{{\bm{a}}}
\def\vb{{\bm{b}}}

\def\ve{{\bm{e}}}

\def\vh{{\bm{h}}}

\def\vu{{\bm{u}}}
\def\vv{{\bm{v}}}

\def\mA{{\bm{A}}}

\def\mI{{\bm{I}}}

\def\mR{{\bm{R}}}

\def\mU{{\bm{U}}}

\DeclareMathAlphabet{\mathsfit}{\encodingdefault}{\sfdefault}{m}{sl}
\SetMathAlphabet{\mathsfit}{bold}{\encodingdefault}{\sfdefault}{bx}{n}

\def\gG{{\mathcal{G}}}

\def\gN{{\mathcal{N}}}

\def\gS{{\mathcal{S}}}
\def\gT{{\mathcal{T}}}

\usepackage{ mathdots  }
\usepackage{ amsmath   }
\usepackage{ amssymb   }
\usepackage{ color     }
\usepackage{ graphicx  }
\usepackage{ amsfonts  }
\usepackage{ epstopdf  }
\usepackage{ upgreek   }
\usepackage{ bm        }
\usepackage{ graphicx  }
\usepackage{ subfigure }
\usepackage{ mathtools }
\usepackage{ algorithm }
\usepackage{algorithmic}
\usepackage{ booktabs  }
\usepackage{ multirow  }
\usepackage{ caption   }
\usepackage{ array     }
\usepackage{ float     }
\usepackage{   soul    }
\usepackage[normalem]{ulem}

\AtBeginDocument{%
	\providecommand\BibTeX{{%
			\normalfont B\kern-0.5em{\scshape i\kern-0.25em b}\kern-0.8em\TeX}}}

\acmConference[DLP-KDD 2022]{4th Workshop on Deep Learning Practice and Theory for High-Dimensional Sparse and Imbalanced Data with KDD 2022}{August 14, 2022}{Virtual Conference, Online}
\acmPrice{15.00}
\acmISBN{978-1-4503-XXXX-X/22/08}

\title{Flattened Graph Convolutional Networks For Recommendation}
\begin{document}
	\author{Yue Xu}
	\email{yuexu.xy@foxmail.com}
	\affiliation{%
		\institution{Alibaba Group.}
		\state{}
		\country{}
	}
	
	\author{Hao Chen}
	\email{sundaychenhao@gmail.com}
	\affiliation{%
		\institution{Tencent Inc.}
		\state{}
		\country{}
	}
	
	\author{Zengde Deng}
	\email{dengzengde@gmail.com}
	\affiliation{%
		\institution{Cainiao Network.}
		\state{}
		\country{}}
	
	\author{Yuanchen Bei}
	\email{yuanchenbei@zju.edu.cn}
	\affiliation{%
		\institution{Zhejiang University.}
		\state{}
		\country{}}
	
	\author{Feiran Huang}
	\email{huangfr@jnu.edu.cn}
	\affiliation{%
		\institution{Jinan University.}
		\state{}
		\country{}}
	
\begin{abstract}
Graph Convolutional Networks (GCNs) and their variants have achieved significant performances on various recommendation tasks. However, many existing GCN models tend to perform recursive aggregations among all related nodes, which can arise severe computational burden to hinder their application to large-scale recommendation tasks. To this end, this paper proposes the flattened GCN~(FlatGCN) model, which is able to achieve superior performance with remarkably less complexity compared with existing models. Our main contribution is three-fold. First, we propose a simplified but powerful GCN architecture which aggregates the neighborhood information using one flattened GCN layer, instead of recursively. The aggregation step in FlatGCN is parameter-free such that it can be pre-computed with parallel computation to save memory and computational cost. Second, we propose an informative neighbor-infomax sampling method to select the most valuable neighbors by measuring the correlation among neighboring nodes based on a principled metric. Third, we propose a layer ensemble technique which improves the expressiveness of the learned representations by assembling the layer-wise neighborhood representations at the final layer. Extensive experiments on three datasets verify that our proposed model outperforms existing GCN models considerably and yields up to a few orders of magnitude speedup in training efficiency.
\end{abstract}
\maketitle


\section{Introduction}
Graph Convolutional Networks (GCNs), which generalize the Convolutional Neural Networks~(CNNs) on graph-structured data~\cite{kipf2016semi}, have achieved impressive performance on various graph-based learning tasks~\cite{velivckovic2017gat,hamilton2017representation,hamilton2017graphsage}, including recommendation~\cite{zhao2019intentgc}. 
The core idea behind GCNs is to iteratively aggregate information from locally nearby neighbors in a graph using neural networks~\cite{chen2018fastgcn}. Each node at one GCN layer performs graph convolution operations to aggregate information from its nearby neighbors at the previous layer. 
By stacking multiple GCN layers, the information can be propagated across the far reaches of a graph, which makes GCNs capable of learning from both content information as well as the graph structure. 
As such, GCN-based models are widely adopted in recommendation tasks~\cite{ying2018graph,zhao2019intentgc, fan2019meirec, wu2019dual,  chen2022neighbor,wang2019knowledge, Wang2019graph,chen2021non,chen2022generative} which require learning from relational datasets. 
However, although existing GCN-based recommendation models have set new standards on various benchmark tasks~\cite{ying2018graph,zhao2019intentgc, fan2019meirec,wang2019kgat, wang2019knowledge, chen2020label, wang2019ngcf,he2020lightgcn,chen2020graph}, many of them suffer from two pitfalls.

First, many existing GCN models suffer from high computational complexity due to the use of multi-layer architectures and complicated modeling techniques. This may largely hinder their application to large-scale real-world recommender systems. 
For example, the metapath-guided GCN models~\cite{hu2018leveraging, fan2019meirec} propose to construct manifold metapaths for neighborhood aggregation, which introduces high complexity on data pre-processing and information aggregation. Meanwhile, the attention-based GCN~(GAT) models propose to generalize graph convolution with the attention mechanism~\cite{velivckovic2017gat, wang2019kgat, wu2019dual}, which, however, introduces an excessive amount of additional parameters to fit. 
On the other hand, the recent advances on simplified GCNs such as~\cite{wu2019sgc,he2020lightgcn}, indicate that it is feasible to remove certain components from existing architectures while still maintaining comparable performances. 
This motivates us to rethink the essential components of building an expressive GCN model for recommendations. 

Second, the recursive neighborhood aggregation among all nodes arises severe computational burden but may have limited contribution in recommendation tasks.
Specifically, as pointed out in~\cite{li2018laplacian}, the convolution in the GCN model is indeed a special form of Laplacian smoothing, 
which makes the feature of nodes within the same cluster similar to greatly ease the classification or regression task.
Therefore, \textit{it is critical for GCN models to ensure that similar nodes have been grouped into the same cluster before performing neighborhood aggregations.}
In homogeneous networks, it is highly likely for two similar nodes to form a direct edge~\cite{mcpherson2001homophily}.
However, the networks are heterogeneous in the context of recommendation. As such, the difficulty of recognizing similar nodes arises since we need to measure the correlation between any user-item pair based on their \textit{indirect} relationships. 
Existing models usually measure the correlation by counting the historical interactions~\cite{fan2019meirec, Wang2019graph, wu2019dual, wang2019knowledge}. 
However, such intuitive measurements can be easily dominated by the popular user/items. Besides, the correlations are unlikely to scale linearly with the number of historical interactions. 
Therefore, it is worth exploring the definition of a principled metric to quantitatively evaluate the importance of neighbors in heterogeneous networks. This would also pave the way for the design of an informative and efficient neighbor sampling method for GCN models.


In this paper, we propose the flattened GCN~(FlatGCN) model which has a much lower complexity compared to existing GCN-based recommendation models but is able to achieve superior performance.
The main contributions are summarized as follows.
First, we propose the FlatGCN to simplify the recursive neighborhood aggregation in standard GCNs as flattened layer aggregations which require performing propagation only once over a single GCN layer. Moreover, the aggregation step in FlatGCN is a parameter-free operation such that it can be pre-computed to save vast memory and computational cost thereby easing the training on large-scale graphs when applied to real-world recommender systems. 
Second, we propose an informative and efficient sampling method named neighbor-infomax to select the top informative neighbors for neighborhood aggregation. Specifically, we propose a principled metric to explicitly measure the correlation among neighboring nodes in a user-item bipartite graph and then select the most informative neighbors according to the evaluation results so as to improve the quality of neighborhood aggregation. 
Third, we propose a layer ensemble technique which improves the expressiveness of the learned representations by assembling the layer-wise neighborhood representations at the final layer. 
Extensive experiments on two benchmarks and one commercial dataset verify that our proposed model outperforms existing GCN models considerably and yields up to a few orders of magnitude speedup in training, in terms of recommendation tasks.

\section{METHODOLOGY}
In this section, we first introduce the neighbor information measurement to rank the neighbor set and select the most informative subset. 
Then we present the efficient and powerful flattened layer aggregation to represent the subgraph of a given user~(item). 
Finally, we propose a novel layer ensemble architecture to predict the relevance score. 
We assume that we have learned users~(items)' embeddings with given collaborative filtering models such as Meta2Vec~\cite{dong2017metapath2vec} or LightGCN~\cite{he2020lightgcn} 
which are set as the node features for the users~(items).

\begin{figure}[tb]
	
	\centering
	\includegraphics[trim = 20 10 10 10, clip, width=0.85\columnwidth]{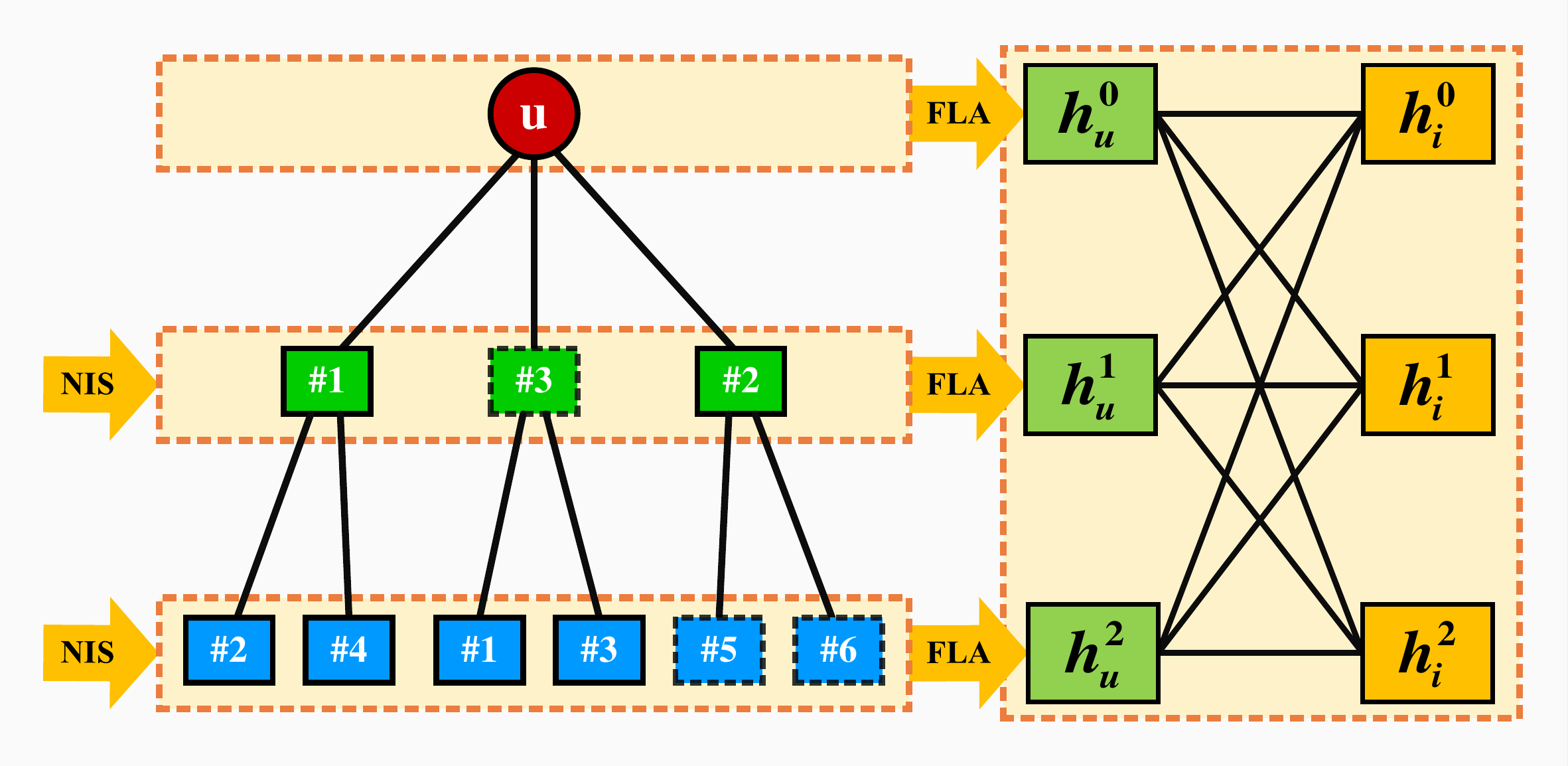}
	\vskip -1.0em
	\caption{Sketch of FlatGCN. 
		On the left, we present the flattened layer aggregation, utilizing the neighbor information measurement to rank the neighbors and select the top informative neighbors from each layer's neighbor set.
		For example, the sampling size of the middle layer is 2, and we select the top-2 informative neighbors, while the sampling size for the bottom layer is 4, and we select top-4 informative neighbors. 
		The layer ensemble architecture is shown on the right.}
	\label{fig:framework}
	\vskip -1.5em
\end{figure}

\subsection{Neighbor Information Measurement}
Starting from GraphSAGE~\cite{hamilton2017graphsage} and FastGCN~\cite{chen2018fastgcn}, GCN models downsample the neighbors to accelerate the recursive message passing. 
Since this downsampling process would inevitably lose information, it is critical to preserve as much information as possible in the preserved neighbor, which could maximally represent the root node. 
To this end, we introduce the mutual information measurement to evaluate to what extent a neighbor can describe the given root node. 
Besides, we also propose a simplified version of neighbor-root-node information measurement for efficient computation. 


\smallskip\noindent\textbf{Neighbor-Root-Node Mutual Information.} 
Let $u$ denote the root node. 
We first select an arbitrary node $v\in \gN_u$. 
Let the random variable $\vv$  be the feature of a randomly picked node in $\gN_u$, then the distribution of $\vv$ is $P_\vv = P(\vv = \ve_v)$, where $\ve_v$ is the outcome feature of node $v$. 
Similarly, we assume $\vu$ be the feature associated with the root node $u$, then the distribution of $\vu$ is $P_\vu = P(\vu = \ve_u)$,  where $\ve_u$ is the outcome feature of node $u$. 
After defining these notations, their mutual information $I(\vv,\vu)$ is the KL-divergence between the joint distribution $P_{\vv,\vu} = P(\vv = \ve_v, \vu = \ve_u)$ and the product of marginal distributions $P_\vv \otimes P_\vu$:
\begin{small}
	\begin{equation}
		\setlength{\abovedisplayskip}{3pt} 
		\setlength{\belowdisplayskip}{2pt}
		\begin{split}
			& I(\vv,\vu) = D_\text{KL}(P_{\vv,\vu}\parallel P_\vv \otimes P_\vu)\\
			\overset{(a)}{\geq} &\sup\limits_{T\in \gT} \{ \mathbb{E}_{\ve_v,\ve_u\thicksim P_{\vv,\vu}}[T(\ve_v,\ve_u)] \\
			& - \mathbb{E}_{\ve_v\thicksim P_{\vv,\vu'}\thicksim P_{\vu'}}[e^{T(\ve_v, \ve_{u'})-1}]
			\},
		\end{split}
		\label{eq:kl}
	\end{equation}
\end{small}
where $(a)$ follows from $f$-divergence representation based on KL-divergence~\cite{mohamed2018mutual}; random variable $\vu'$ denotes the feature associate with an arbitrary node $u'\in \mU\cup\mI$; $T\in\gT$ is an arbitrary function that maps a pair of features to a real value, which reflecting the correlation of two features. 
After replacing the $f$-divergence in Eq.~(\ref{eq:kl}) with a GAN-like divergence, Eq.~(\ref{eq:kl}) is written as follows,
\begin{small}
	\begin{equation}
		\setlength{\abovedisplayskip}{3pt} 
		\setlength{\belowdisplayskip}{2pt}
		\begin{split}
			I_{\text{GAN}}(\vv,\vu) \geq &\sup\limits_{T\in \gT}\{
			\mathbb{E}_{P_{\vv,\vu}}[\text{log}\sigma(T(\ve_v,\ve_u))] \\&+ \mathbb{E}_{P_\vv,P_{\vu'}}[\text{log}(1-\sigma(T(\ve_v,\ve_{u'})))]
			\},
		\end{split}
		\label{eq:gan}
	\end{equation}
\end{small}
where $\sigma(\cdot)$ is the sigmoid function. 

\noindent\textbf{Practical Simplification.} 
In practice, we cannot go over the entire functional space to evaluate the exact value of $I_{\text{GAN}}(\vv,\vu)$. 
Since the feature of nodes are learned collaborative filtering embeddings, we parameterize $T(\va,\vb)$ by computing their inner product as $T(\va,\vb) = \va^\top \cdot \vb$. 
Through the trained CF embeddings, we obtain a lower bound of GAN-based mutual information as:
\begin{small}
	\begin{equation} 
		\setlength{\abovedisplayskip}{3pt} 
		\setlength{\belowdisplayskip}{2pt}
		\begin{split}
			I_{CF}(\vv,\vu) = & \log\sigma(\ve_v^\top\cdot\ve_u) \\  & +\frac{1}{N+M}\sum_{u'\in \mU\cup\mI}\log(1-\sigma(\ve_v^\top \cdot \ve_{u'})),
		\end{split}
		\label{eq:cri}
	\end{equation}
\end{small}
where $N=|\mU|$ and $M=|\mI|$ denote the numbers of users and items, respectively. 
In $I_{CF}(\vv,\vu)$, the first term reflects the correlation between node $v$ and the root node $u$, meaning to what extend does the node $v$ describe the root node $u$. 
The second term evaluates the difference between the sampling set and the whole user-item embeddings, which estimates the particularity of the node $v$. 
Since $\log(1-\sigma(\cdot))$ is a concave function, we can further simplify the second term of Eq.~(\ref{eq:cri}) with Jensen's inequality to obtain the following neighbor information evaluation function,
\begin{small}
	\begin{equation}
		\setlength{\abovedisplayskip}{3pt} 
		\setlength{\belowdisplayskip}{2pt}
		C(v,u) = \log\sigma(\ve_v^\top\cdot \ve_u) + \log(1-\sigma(\ve_v^\top\cdot \overline{\ve})),
		\label{eq:nie}
	\end{equation}
\end{small}
where $\overline{\ve} = 1/(N+M)\sum_{u'\in{\mU\cup\mI}} \ve_{u'}$ denotes the average embedding of all the user-item embeddings. 
With this neighbor information evaluation function Eq.~(\ref{eq:nie}), we propose the neighbor infomax sampling and the flattened layer aggregation in the next subsection.

\subsection{Flattened Layer Aggregation}
Mathematically, let $\gG(\mU,\mI,\mR)$ denote the user-item interaction graph. $\mU=\{u_1,\cdots,u_N\}$ denotes the user set, and $\mI= \{i_1,\cdots,i_M\}$ denotes the item set, while $\mR\in\mathbb{R}^{N\times M}$ describes the user-item interaction matrix. 
After introducing the notations, the argument adjacent matrix for users and items can be defined as follows,
\begin{small}
	\begin{equation}
		\setlength{\abovedisplayskip}{3pt} 
		\setlength{\belowdisplayskip}{2pt}
		\mA = 
		\left[
		\begin{matrix}
			\textbf{0} & \mR  \\
			\mR^\top & \textbf{0} 
		\end{matrix}
		\right] \in \mathbb{R}^{(N+M)\times (N+M)}.
	\end{equation}
\end{small}
Therefore, we can define the multi-hop neighbors of a given node $u$ according to the adjacent matrix $\mA$. 
Specially, $\gN_u^0=\{u\}$. 
The first-order neighbor set is given by $\gN_u^1=\{v|\mA_{uv} =1\}$, while the second-order neighbor set is denoted by $\gN_u^2=\{v|\mA_{uv}^2 \geq1\}$, so on and so forth. 

As shown in Fig.~\ref{fig:framework}, for a given root node $u$, we select the most informative subset $\gS_u^k$ from a given neighbor set $\gN_u^k$ by greedily selecting the top-$S^k$ informative neighbors, namely,
\begin{small}
	\begin{equation}
		\setlength{\abovedisplayskip}{3pt} 
		\setlength{\belowdisplayskip}{2pt}
		\gS_u^k = \text{top-rank}\{C(v,u)\ \text{for\ } v \in \gN_u^k, \ S^k\},
		\label{eq:top}
	\end{equation}
\end{small}
where top-rank is the function that returns the indices of the top-$S^k$ value and $S^k$ denotes the sampling size of the $k$-th layer. 
Based on the definition of the neighbor infomax selection in Eq.~(\ref{eq:top}), we can utilize flat infomax aggregation to extract the information of different layers,
\begin{small}
	\begin{equation}
		\setlength{\abovedisplayskip}{3pt} 
		\setlength{\belowdisplayskip}{2pt}
		\vh_u^k = \frac{1}{S^k}\sum_{v \in \gS_u^k} \ve_v.
		\label{eq:agg}
	\end{equation}
\end{small}
Noted that the aggregation step~Eq.~(\ref{eq:top}) and Eq.~(\ref{eq:agg}) is a parameter-free operation such that it can be done in a pre-processing manner only once, thereby saving vast memory and computational cost to ease the training on large-scale graphs and application to real-world recommender systems. 

\subsection{Layer Ensemble Architecture}
With Eq.~(\ref{eq:agg}), FlatGCN aggregates the information of a $K$-layer subgraph into $K$ vectors. 
Thus for any give user-item pair $(u,i)$, we have $\{\vh_u^0,\cdots,\vh_u^K\}$ to represent the user subgraph and $\{\vh_i^0,\cdots,\vh_i^K\}$ to represent the item subgraph.

The success of  FM~\cite{guo2017deepfm,juan2016ffm} inspires that a hand-crafted cross-interaction function may be better than brutally learning with Multi-Layer Perception~(MLP) function. 
As shown in Fig.~\ref{fig:framework}, we first compute the interactions between different layers combinations for every layer of user's and the item's subgraphs. 
In such a way, we can easily take full advantage of the aggregation embedding of each subgraph layer for both user and item, and thus we model the complex interaction from user $u$ to item $v$ as,
\begin{small}
	\begin{equation}
		\setlength{\abovedisplayskip}{3pt} 
		\setlength{\belowdisplayskip}{2pt}
		\hat{y}_{uv} = \text{MLP}[\parallel_{i=0}^K\parallel_{j=0}^K(\vh_u^i)^\top\cdot\vh_v^j],
		\label{eq:ensemble}
	\end{equation}
\end{small}
where $\parallel$ indicates the concatenation operation, $(\vh_u^i)^\top\cdot\vh_v^j$ denotes the relevance score between the $i$th-layer representation of the user subgraph and $j$th-layer of the item subgraph. 
It is noteworthy that Eq.~(\ref{eq:ensemble}) actually summarize the complicated relations between the user-item pair $(u,i)$ into a $K^2$-dimension vectors, that greatly reduce the number of the parameters in the followed MLP function.

\label{sec:result}
\begin{table*}[ht]
	\small
	\centering
	\caption{Overall Performance Comparison}
	\resizebox{\textwidth}{!}{
		\setlength{\tabcolsep}{2.7mm}{
			\begin{tabular}{cc|ccc|ccc|ccc}
				\toprule
				& \multicolumn{1}{c}{} & \multicolumn{3}{c}{\textbf{AmazonBook}} & \multicolumn{3}{c}{\textbf{Yelp2018}} & \multicolumn{3}{c}{\textbf{WeChat}} \\
				\midrule
				Embeddings  & Methods & PRE   & REC   & NDCG  & PRE   & REC   & NDCG  & PRE   & REC   & NDCG \\
				\midrule
				\multirow{6}[4]{*}{Meta2Vec} & IntentGC & 0.0259 & 0.0517 & 0.0458 & 0.0274 & 0.0592 & 0.0490 & 0.0116 & 0.0908 & 0.0512 \\
				& GCN   & 0.0271 & 0.0543 & 0.0476 & 0.0299 & 0.0648 & 0.0527 & 0.0181 & 0.1425 & 0.0789 \\
				& NGCF  & 0.0384 & 0.0769 & 0.0678 & 0.0363 & 0.0769 & 0.0666 & 0.0189 & 0.1493 & 0.0837 \\
				& LightGCN & \underline{0.0448} & \underline{0.0919} & \underline{0.0833} & \underline{0.0388} & \underline{0.0842} & \underline{0.0704} & \underline{0.0191} & \underline{0.1478} & \underline{0.0940} \\
				& FlatGCN & \textbf{0.0498} & \textbf{0.1014} & \textbf{0.0918} & \textbf{0.0430} & \textbf{0.0932} & \textbf{0.0807} & \textbf{0.0258} & \textbf{0.2019} & \textbf{0.1233} \\
				\cmidrule{2-11}      & Improve & 11.16\%* & 10.34\%* & 10.20\%* & 10.82\%* & 10.69\%* & 14.63\%* & 35.08\%* & 35.23\%* & 31.17\%* \\
				\midrule
				\multirow{6}[4]{*}{LightGCN} & IntentGC & 0.0307 & 0.0609 & 0.0540 & 0.0335 & 0.0723 & 0.0616 & 0.0230 & 0.1835 & 0.1057 \\
				& GCN   & 0.0317 & 0.0649 & 0.0576 & 0.0291 & 0.0625 & 0.0529 & 0.0233 & 0.1872 & 0.1079 \\
				& NGCF  & 0.0433 & 0.0880 & 0.0785 & 0.0401 & 0.0853 & 0.0737 & \underline{0.0425} & \underline{0.3533} & \underline{0.2015} \\
				& LightGCN & \underline{0.0450} & \underline{0.0919} & \underline{0.0823} & \underline{0.0410} & \underline{0.0874} & \underline{0.0752} & 0.0342 & 0.2831 & 0.1671 \\
				& FlatGCN & \textbf{0.0485} & \textbf{0.0982} & \textbf{0.0884} & \textbf{0.0450} & \textbf{0.0954} & \textbf{0.0818} & \textbf{0.0709} & \textbf{0.5932} & \textbf{0.3543} \\
				\cmidrule{2-11}      & Improve & 7.78\%* & 6.86\%* & 7.41\%* & 9.76\%* & 9.15\%* & 8.78\%* & 66.82\%* & 67.90\%* & 75.83\%* \\
				\bottomrule
			\end{tabular}%
	}}
	\label{tab:result}
\end{table*}

\section{Experiments}
\subsection{Experiment Setup}

\noindent\textbf{Datasets.}
We utilize two benchmark datasets and a commercial dataset to evaluate the recommendation performance:
AmazonBook and Yelp2018 are two benchmark datasets used in~\cite{wang2019ngcf,he2020lightgcn}.
The WeChat dataset contains users' clicks on different articles recorded by the WeChat platform.
We use Meta2Vec~\cite{dong2017metapath2vec} and the state-of-the-art~(SOTA) LightGCN~\cite{he2020lightgcn} to produce the pre-trained embeddings of users and items in datasets to feed into the GCN model as the raw features.

\noindent\textbf{Evaluation Protocols.}
We randomly split the entire user-item recommendation records of each dataset into an embedding training set, GCN model training~(validation) set, and a test set to evaluate the performance, where each of them contains 65\%, 15\%, and 20\% of the full records, respectively. 
Besides, we apply the wildly used full-rank evaluation protocols~\cite{he2020lightgcn} to evaluate the recommendation performance.

\noindent\textbf{Comparison Methods.}
We compare the following SOTA baselines: 
\textbf{I.}~GCN~\cite{hamilton2017graphsage} applies 2-layer GCN to process input embeddings following PinSAGE~\cite{ying2018graph}, followed by an inner product function.
\textbf{II.}~IntentGC~\cite{zhao2019intentgc} introduces a fast architecture named IntentNet to avoid unnecessary feature interactions for efficient training. \textbf{III.}~NGCF*~\cite{wang2019ngcf} is an embedding learning method. We fix its inputs as the given user-item embeddings and only train its propagation layers. We denote it as NGCF*.
\textbf{IV.}~LightGCN*~\cite{he2020lightgcn} is the SOTA recommendation model. Similar to NGCF*, we fix its inputs as the given user-item embeddings and only train its propagation layers.

\noindent\textbf{Parameter Settings.}
The optimal parameter settings for all comparison methods are achieved by either empirical study or suggested settings by the original papers. We consider the 3-layer subgraphs and set the sampling size $S_k=25$ for $k=1,2$. Especially, we adopt Adam~\cite{kingma2014adam} as the optimizer with learning rate as $0.001$; the $L_2$ regularization weight as $10^{-5}$.

\begin{table}[t]
	\centering
	\caption{Comparison of sampling policies.}
	\resizebox{1\columnwidth}{!}{
		\setlength{\tabcolsep}{2.5mm}{
			\begin{tabular}{ccccc}
				\toprule
				Embeddings & Sampling & WeChat(PRE) & WeChat(REC) & WeChat(NDCG) \\
				\midrule
				\multirow{4}[4]{*}{Meta2Vec} & Random & 0.0237 & 0.1858 & 0.1104 \\
				& Intuitive & \underline{0.0241} & \underline{0.1892} & \underline{0.1115} \\
				& Infomax & \textbf{0.0258} & \textbf{0.2019} & \textbf{0.1233} \\
				\cmidrule{2-5}      & Improve & 7.05\% & 6.71\% & 10.58\% \\
				\midrule
				\multirow{4}[4]{*}{LightGCN} & Random & 0.0659 & 0.5547 & 0.3268 \\
				& Intuitive & \underline{0.0685} & \underline{0.5750} & \underline{0.3388} \\
				& Infomax & \textbf{0.0709} & \textbf{0.5932} & \textbf{0.3543} \\
				\cmidrule{2-5}      & Improve & 3.50\%* & 3.17\%* & 4.57\%* \\
				\bottomrule
			\end{tabular}%
	}}
	\label{tab:samp}
\end{table}

\subsection{Result Analysis}
\noindent\textbf{Performance Comparison.}
Table~\ref{tab:result} reports the performance on the three datasets w.r.t. precision~(PRE@20), recall~(REC@20), and ndcg~(NCDG@20) metrics. 
The improvement gives the relative enhancement of FlatGCN over the best baseline~(underlined). The superscript * denotes statistical significance against the best baseline with $p<0.05$.

Overall, our proposed FlatGCN statically significantly outperforms the best base baselines among all three datasets w.r.t. all evaluation metrics and both Meta2Vec and LightGCN embeddings. These results verify the consistent superiority of FlatGCN on characterizing the relevance between users and items than SOTA GCN-based recommendation methods.
Particularly for the commercial WeChat dataset, FlatGCN leads to an enhanced performance by more than 30\% on Meta2Vec embeddings. With LightGCN embeddings, FlatGCN outperforms the best baselines by over 60\%.  

\begin{figure}[t]
	\centering	
	\includegraphics[trim = 1 5 1 1, clip, width=0.8\columnwidth]{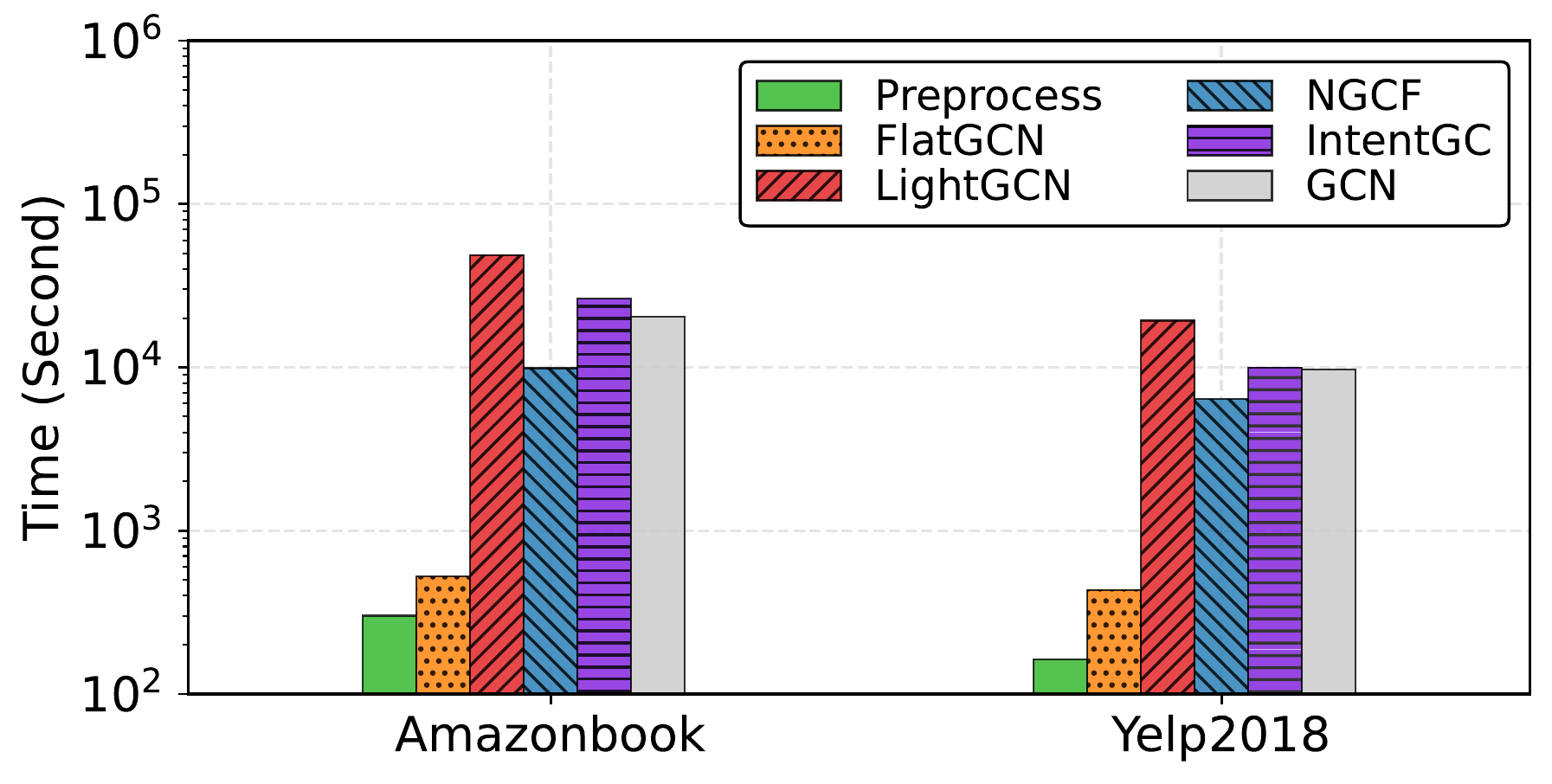}
	\caption{Comparison of the training time. All methods are running on the same GPU device.}
	\vskip -2em
	\label{fig:eff}
\end{figure}

\smallskip\noindent\textbf{Learning Efficiency.}
One main advantage of FlatGCN is the low training complexity. In Fig.~\ref{fig:eff}, we present the average training time of FlatGCN and the other baselines. 
We also present the preprocessing time of FlatGCN. 
The results in Fig.~\ref{fig:eff} shows that FlatGCN achieves superior performance with one or two orders of magnitude speedup in training time. 
Notably, FlatGCN is $10$ times faster than the fastest baseline: NGCF. Moreover, the preprocessing time of FlatGCN can be neglected, comparing with the training time of the baselines.

\smallskip\noindent\textbf{Sampling Comparison.}
In~\autoref{tab:samp}, we compare three different neighbor sampling methods on the WeChat dataset:
1)~\textit{Random:}~Random walk-based sampling~\cite{ying2018graph}, which simulates random walks starting from each node and computes the L1-normalized visit count of neighbors visited by the random walk. 
2)~\textit{Intuitive:}~First-order proximity-based sampling~\cite{fan2019meirec, Wang2019graph, wu2019dual}, examining the neighborhood similarity based on the edge weights~(e.g., number of clicks). 
3)~\textit{Infomax:}~Our proposed neighbor infomax selection.
Overall, the random sampling method generates the worst performance. Intuitive sampling outperforms random sampling since the edge weights can represent the importance of an edge. Infomax sampling method achieves the best performance, due to its ability to select the most informative subset from the neighbor set.

\section{Conclusion}	
In this paper, we introduced the FlatGCN model which is able to achieve superior performance along with a few orders of magnitude speedup in training compared with existing models.
Specifically, we proposed a flattened GCN architecture which is able to perform neighborhood propagation only once over a single GCN layer instead of recursively. Moreover, we proposed an informative sampling method named neighbor-infomax to select the top informative neighbors for neighborhood aggregation. Finally, we proposed a layer ensemble technique to improve the model expressiveness. 
Extensive experiments verified the superiority of the proposed FlatGCN on both learning performance and training speed.

\bibliographystyle{unsrt}
\bibliography{references}
	
\end{document}